# Periodic Collinear-Slotted Leaky Wave Antenna With Open Stopband Elimination

Alireza Mallahzadeh, *Senior Member, IEEE*, and Sajad Mohammad-Ali-Nezhad, *Member, IEEE*

*Abstract*—A low cross-polarization periodic-slotted ridged substrate-integrated waveguide (RSIW) leaky wave antenna (LWA) array is presented. Parametric constraints of the structure were investigated to realize a periodic leaky wave antenna (PLWA) with the ability of scanning from backward endfire into the forward quadrant. An appropriate multimode transverse equivalent network (TEN) is presented for the proposed antenna and values of leakage rate and phase constant for different parameters of the structure were extracted using the transverse resonance technique (TRT). The obtained results were compared to those achieved from HFSS software simulations. The open stopband in the broadside is eliminated using certain dimensions for the structure. By simultaneously manipulating a number of parameters, a variable $\alpha$ and constant $\beta$ are resulted so that the desired sidelobe level (SLL) can be realized. The proposed structure was simulated and then manufactured. Simulation results show good agreement with measurement results.

*Index Terms*—Leaky wave antenna (LWA), low cross-polarization, open stopband elimination and slot array antenna, ridged substrate waveguide, transverse resonance technique (TRT).

## I. INTRODUCTION

LONGITUDINALLY slotted waveguide array antennas have various applications in radar, communications, and navigation systems because of their high gain and efficiency, low loss, and the possibility to easily control their radiation patterns [1]–[4].

Slotted waveguide array antennas are categorized into two main groups, including the standing wave antennas and traveling wave antennas. In the standing wave mode, the slots have resonance lengths and are fed in an in-phase manner, which results in an antenna structure radiating in the broadside direction when used in an array antenna structure. However, for having radiating slots, they need to have an offset from the centerline of the upper plane of the waveguide. In this situation, the slots have a distance of $\lambda_g/2$ from their adjacent slots and are alternatively placed on both sides of the centerline, which results in an increase in the cross-polarization and creation of second-order beams [2]–[6]. In the traveling wave mode, the slots do not require resonance lengths and in-phase feed. Consequently, the distance between the slots and slots lengths does not depend on the wavelength, resulting in structures with higher impedance bandwidths compared to the standing wave structures. Regarding the distance between the slots, changes in frequency cause changes in phase difference between the radiated powers from the slots, resulting in a frequency scanning beam. As the slots are not required to have an in-phase feed in these structures, they do not need to be alternatively placed on both sides of the centerline of the waveguide. In the described structure, second-order beams are not usually formed, but because the slots are offset from the centerline, cross-polarization is increased [7]–[9].

In recent years, slotted substrate-integrated waveguide (SIW) array antenna structures have attracted considerable attention [10]–[15].

Leaky wave antennas (LWAs) are among the different types of traveling wave antennas. LWAs can be categorized into two main types based on their geometry and principle of operation. The first type is the uniform LWA, which is uniform along the length of the guiding structure. In this type, the dominant mode is a fast wave and the maximum value of the antenna radiation pattern is in the forward quadrant [16]–[27].

The second type is the periodic LWA in which the dominant mode is a slow wave and cannot radiate. Introduction of the periodic array as discontinuities generates an infinite number of space harmonics (SHs). Some of these SHs are fast waves. Actually, it is the periodic modulation that makes radiation from the structure possible. The scan range in this type of LWAs is from backward quadrant into the forward quadrant. However, open stopbands usually occur around the broadside [28]–[34].

In this paper, a low cross-polarization collinear-slotted periodic leaky wave array (LWA) antenna based on a double-ridged substrate-integrated waveguide (RSIW) structure is presented. As this structure is a slow wave and the slots, i.e., the discontinuities are placed only on the centerline without any displacement and also because radiation can take place through changing the characteristics of the basic RSIW structure, it can be analyzed using methods used in periodic leaky wave antennas (PLWAs). Moreover, cross-polarization in these structures is reduced and, i.e., because the slots do not have an offset from the centerline.

One of the methods for analyzing LWA structures is based on the transverse equivalent network (TEN) [35]–[43]. In this paper, the leaky wave propagation constant of the proposed structure is obtained from the resonance of the TEN using a multimode and accurate TEN method. In Section II, the principle of operation of the proposed PLWA is presented. In Section III, the constraints that the proposed PLWA structure needs to have so that only a single main lobe is created when scanning from the backward quadrant into the forward quadrant are discussed. The proposed TEN is presented in Section IV. In Section V, simulation results, using high frequency structure simulator (HFSS) software, for the proposed structure, are presented and the leakage rate and phase constant obtained from simulations are compared with the leakage rate and phase constant achieved from the multimode TEN that showed a good agreement. In Section VI, for achieving the desired sidelobe level (SLL), a graph of the dispersion relation is presented. The proposed LWA is designed and manufactured based on this graph and Chebyshev coefficients for SLL = −25 dB.

## II. Principle of Operation of the Slotted RSIW Array LWA

If the dominant mode in the SIW structure is slow wave, it is possible to use the SIW as the basic structure of the PLWA. In this situation, SHs can be activated through creating discontinuities or periodic modulations in the structure. SHs are fast wave and are capable of radiating. For this purpose, an array of slots can be formed on the upper plane of the SIW.

The field distribution in the slots has to be controlled for having a radiation pattern with a proper SLL. For this reason, the electric field inside the slots is controlled by changing their offset from the centerline. In this case, for obtaining the desired pattern, the basic SIW structure remains intact and the discontinuities are changed. Consequently, the structure cannot be categorized as a PLWA structure. In the proposed structure, two ridges are placed inside the structure, and slots positions are considered to be fixed and the distribution inside the slots can be controlled by changing the dimensions of the ridges. In this situation, discontinuity elements, i.e., the slots, are fixed and the basic RSIW structure is changing and the dispersion relation depends on the dimensions of the structure.

On the other hand, for having radiating slots, they need to have an offset from the centerline of a symmetric SIW structure. This results in an increase in the cross-polarization of the structure and also creation of second-order beams. These structures possess a relatively poor cross-polarization of about −25 dB, whereas the proposed structure has a better cross-polarization and there is no second-order beam in the radiation pattern.

A dissymmetry in the location of the slot, regarding the field inside the waveguide, is needed for having radiating slots. When a slot is placed on the centerline of the upper plane of the waveguide, the electric field on its both sides is symmetrical; therefore, the field must be made asymmetric in any proper possible manner. The ridges inside the waveguide are used for this purpose. The single-element slot antenna with side ridges has a lower cross-polarization [35]; moreover, by placing the slots on the centerline and consequently creating symmetry among array elements, cross-polarization is reduced.

Fig. 1 shows some parts of the proposed antenna. In Fig. 1(a), two substrates with a height of $h$ and a dielectric constant of $\varepsilon_r$ are shown separately. Slots, with a length of $Ls$ and a period of $P$, are located on the centerline of the upper substrate. $a$ shows the distance between the vias on the upper substrate. The lower substrate is used for realizing the ridges. The ridges are located on both sides of the slots and $sr$ is the distance between the two ridges, i.e., the width of the lower SIW. In addition, similar to the upper substrate, $h$ is the thickness of the lower substrate. Cross-section of the proposed antenna is shown in Fig. 1(b), which consists of the two substrates, a slot on the upper substrate and the ridges in the lower substrate. To realize a ridge, a metallic layer consisting of two walls of vias on its sides needs to be formed. As no power can leak from the inner line of vias, i.e., from the inner wall of the ridge, the outer wall of vias is not needed. On each section of the proposed structure, $rw$ shows the width of one of the ridges that equals the center-to-center distance between the vias on the upper and lower substrates and the width of the other ridge equals $a - (sr + rw)$. In Fig. 1(a), the dashed line on the lower substrate corresponds to the center of the vias of the upper substrate; therefore, $rw$ is the distance between the dashed line and the center of the vias of the lower substrate.

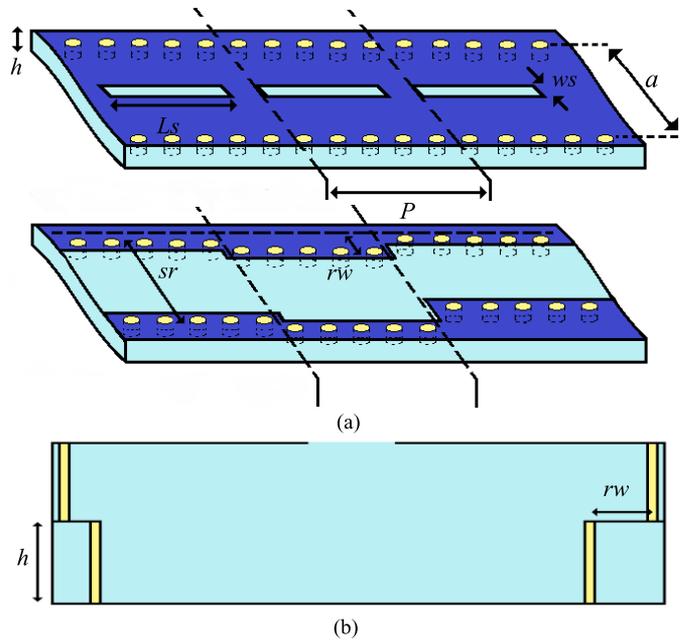

Fig. 1. Collinear slot array RSIW leaky wave antenna. (a) Upper and lower layers. (b) Cross-section.

In Fig. 1, the metallic layers are shown using a dark blue color, the light blue color shows the dielectric layers, and yellow is used for the vias.

## III. Parametric Constraints

To have a single main radiation lobe in the scan range, only the n = −1 SH must be activated in the complete scan range. If higher order modes or other SHs such as the n = −2 SH are generated, there would be more than one main lobe in the radiation pattern. Therefore, the basic RSIW structure must be selected such that only the dominant mode is generated in the whole scanning pattern bandwidth and the n = −1 SH must be the only generated SH and other SHs should not be activated.

Through controlling $\varepsilon_r$, $a$, and $sr$, the proposed structure can be controlled so that only a single mode exists in its bandwidth. If $sr$ has a value close to that of $a$, then it would not have a considerable influence in creating a single mode structure and can be omitted from the calculations. For controlling the generation of SHs, i.e., avoiding the generation of other SHs except the n = −1 SH, the period of the structure $P$ is another parameter that has to be taken into consideration.

Some parametric constraints can be set to achieve the desired scanning capability in the whole impedance bandwidth. The constraints are as follows:

$$\varepsilon_r \geq 2.9 \quad (1)$$
$$P_{max} \cong a\sqrt{\varepsilon_r - 1} \quad (2)$$

where $P_{max}$ is the maximum possible distance between two adjacent slots so that the n = −1 SH is generated from the frequency at which the slow wave is generated in the dominant mode and all the possible impedance bandwidth is used for scanning the radiation pattern. $P_{min}$ is the minimum period of the structure. The constraint for $P_{max}$ depends on the maximum scan angle, i.e., if the main beam scan angle is from endfire to broadside

$$P_{min} \cong 2a/\sqrt{3} \quad (3)$$

and if the scan angle continues into the forward quadrant, then

$$P_{min} \cong \frac{4a\sqrt{\varepsilon_r}}{2 - \sqrt{3}\varepsilon_r}. \quad (4)$$

Through alternatively displacing the ridges regarding the slots, a phase difference of 180° is achieved, which results in an increase in the scanning range in the forward quadrant. This 180°-phase shift in the phase constant of SH modes is because of the phase reversal between adjacent periodic elements. In this situation, the dominant mode is slow wave from the beginning of the impedance bandwidth; note that, in this case, the beginning of the scanning bandwidth and impedance bandwidth are the same. In addition, the n = −1 SH starts radiating from backward endfire and before the generation of higher modes, the maximum scan angle can reach forward endfire or at least a wider scan bandwidth compared to that of the structure without the alternative displacement of the ridges. In the structures with phase reversal between adjacent periodic elements, phase constant is expressed as follows [46]:

$$\beta_n = \beta_0 + (2n + 1)\pi/P \quad (5)$$

where $\beta_n$ is the phase constant of the nth SH and $\beta_0$ is the phase constant of the dominant mode.

For reaching the maximum scan angle in the forward endfire, $P_{min}$ is defined as follows:

$$P_{min} \cong \frac{5a\sqrt{\varepsilon_r}}{2 - \sqrt{3}\varepsilon_r}. \quad (6)$$

In this situation, $P_{max}$ equals (2) and $\varepsilon_r \geq 2.17$ based on $P_{max}$ and $P_{min}$ constraints. In the phase reversal mode, the dielectric constant is less than usual, so the slow wave structure or the PLWA can be created using a lower dielectric constant. In this situation, a higher maximum scan angle can be obtained.

## IV. TRANSVERSE RESONANCE TECHNIQUE

To obtain the dispersion relation, in other words, the phase constant and leakage rate, the transverse resonance technique (TRT) can be used. The first step in using this method is achieving the TEN for the proposed structure.

The slots placed on the upper plane of the proposed RSIW structure, result in the generation of an infinite number of SHs.

For analyzing the slotted RSIW array LWA using the TEN, a period of the structure in a unit cell with phase shift walls is investigated as shown in Fig. 1. Mutual coupling effect is also taken into consideration.

Equivalent network representations of different elements of the structure in the analyzed unit cell have been presented in various previous studies [36]–[41]. In this paper, a complete accurate multimode TEN for the proposed structure is achieved by putting together all of these equivalent network representations for different elements of the antenna.

Fig. 2 shows the proposed TEN as the equivalent circuit for different SHs in the proposed structure.

First, the structure is supposed to be a fast wave, one that can radiate in the dominant mode, and a basic TEN is presented for it. Slots admittances are presented for each SH and then by defining the mutual coupling network, the whole TEN is presented. In the first step, without considering the SHs, i.e., by only considering the common parts of the structure for all modes, the proposed TEN is given based on Fig. 1(b).

For achieving the TEN for each SH mode, the structure must be divided into different parts. Two similar parts are the areas on both sides of the slot. Each area on the left or on the right side of the slot can be modeled as two waveguide components connected to each other; in other words, the area between the slot and the ridge, on the right or the left, is considered as a waveguide that is connected to the next waveguide with a lower height representing the area on top of the ridge. The discontinuity between the two waveguides has to be modeled as well. In the proposed TEN, uniform regions in the waveguide cross-section are represented by transmission lines [42]–[45] and the discontinuities, because of the presence of a ridge on each side of the SIW that results in the creation of two waveguides with different heights on each side, are modeled using a transformer and a lumped element [36]. The discontinuity because of the presence of the slot in the structure is considered in the middle of the proposed TEN and the slot admittance is applied to the structure according to the generated mode.

This part of the proposed TEN is similar to the TEN for a uniform LWA presented in [35] in which the higher order modes are not taken into consideration. In addition, the TEN produced in [35] is used for analyzing a long slot LWA, but the TEN proposed in this paper, because of the generation of SH modes, must be a multimode TEN in which the SHs are taken into consideration.

Inside the RSIW, SHs (with n = ... − 2, −1, 0, 1, 2, 3 ...) are generated, but for having a single main radiation lobe, only the generation of the n = −1 SH is allowed and the radiation impedance characteristics for the slots belong to this mode. These radiation characteristics for a slot with finite length are obtained from [37]. However, other modes

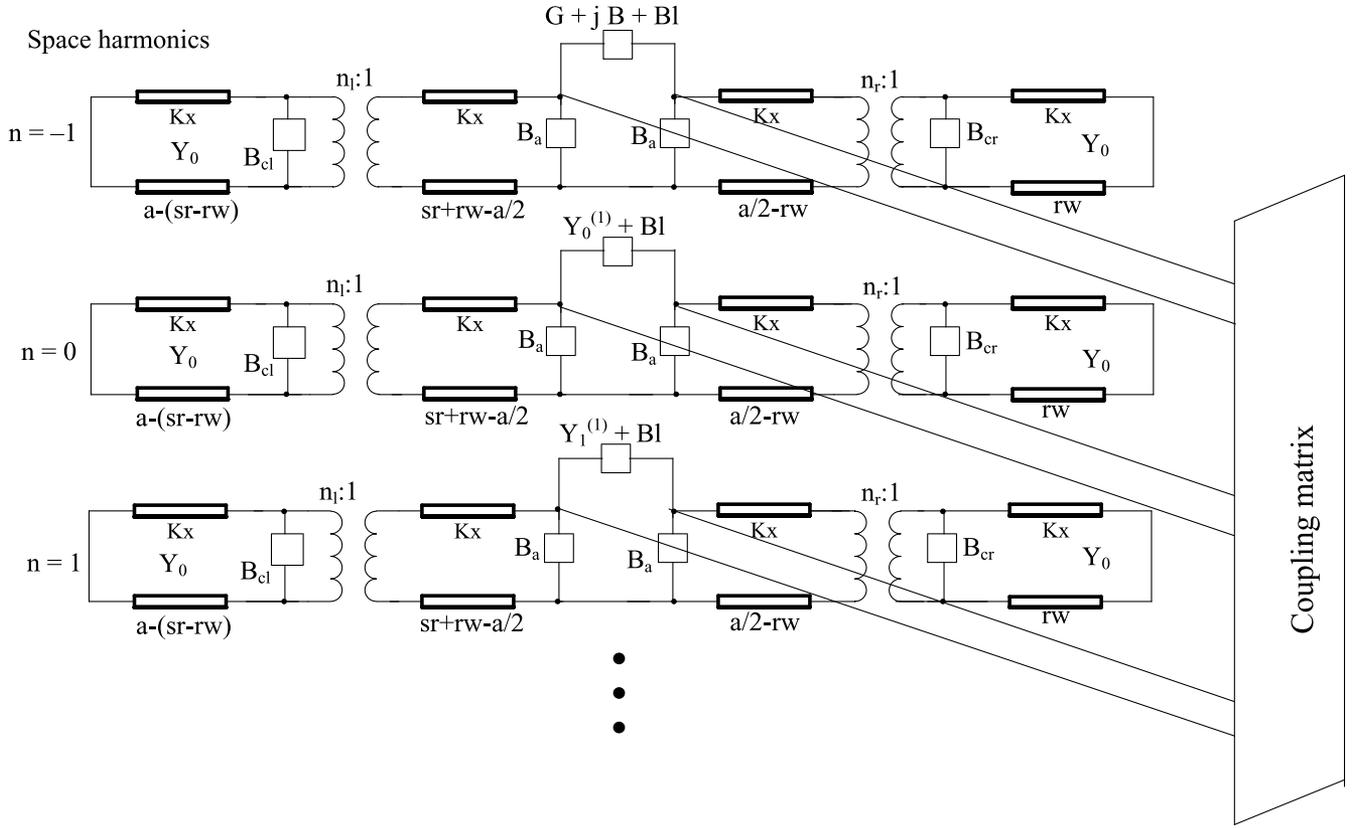

Fig. 2. TEN of the slot array RSIW PLWA.

regarding the air region above the structure are modeled using load admittance ($Y_n^{(1)}$) [28].

The other important part of the proposed TEN is the mutual coupling network for space harmonics. This mutual coupling network is obtained from the calculations in [39] and [40].

When the number of the SH modes increases, the impedance of the coupling circuit is equal to the open circuit and these modes can be neglected. Based on the calculations carried out in this study, the TEN for the proposed structure can produce accurate results considering n = $-2, -1. 0, 1, 2,$ and 3 SHs only and, i.e., because the higher order modes do not effectively influence the results.

The leaky wave propagation constant of the proposed structure can be precisely and effectively calculated through using the achieved TEN in a transverse resonance procedure.

The ordinary TRT is used for achieving the transverse resonance equation of the wavenumber in the proposed structure

$$\vec{Y} + \overleftarrow{Y} = 0 \quad (7)$$

which is applicable to the reference plane just outside the mutual coupling network on port zero.

The TEN indicates that obtaining a simple closed form of analytical expressions is feasible. Consequently, a simple closed-form expression is achieved for the propagation constant of the proposed structure, which makes quick calculation of numerical values possible.

Leakage rate $\alpha$ and phase constant $\beta$ can be calculated through numerically solving the equation resulted from the TRT

and then these parameters can be used for finding the beam angle direction of maximum radiation and beamwidth of the radiation pattern [46]

$$\sin \theta_m \cong \frac{\beta_{-1}}{k_0} \quad (8)$$

$$\Delta \theta \cong \frac{1}{\frac{L_A}{\lambda_0} \cos \theta_m} \approx \frac{\alpha/k_0}{0.183 \cdot \cos \theta_m}. \quad (9)$$

## V. Results for the Propagation Behavior of the Slotted RSIW LWA

Normalized leakage rate and normalized phase constant can be calculated by numerically solving the equation obtained based on the TRT, resulted from the TEN of the proposed slotted RSIW array antenna. $\alpha$ and $\beta$ are also obtained using HFSS software and the method presented in [26].

In this method, leakage rate and phase constant are calculated through sampling the near E-field along the slotted RSIW LWA and then matching this field with the proposed exponential expression based on the wavenumber of the LWA structure.

There are a number of parameters that have effects on $\alpha$ and $\beta$, such as slot length $Ls$, structure period $P$, ridge width $rw$, height of the ridges $h$, and the distance between the two ridges $sr$, i.e., equal to the width of the lower layer of the SIW. Because there are a variety of parameters that affect $\alpha$ and $\beta$, by suitably choosing the values of these parameters, a variable $\alpha$ and fixed $\beta$ can be easily created for achieving the desired radiation pattern. In addition, the open stopband effect can be decreased or even partly removed by properly choosing the values of the structure parameters.

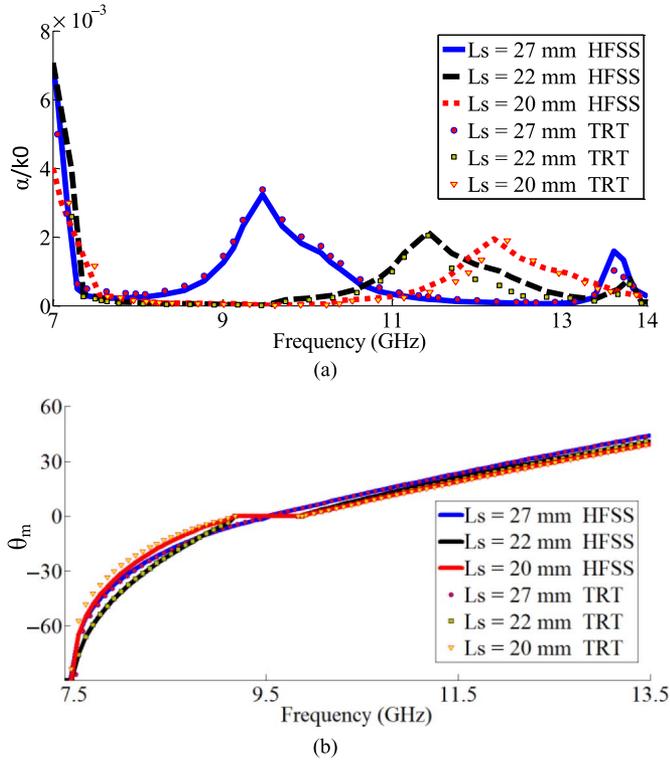

Fig. 3. Simulated complex propagation constant for the proposed structure. (a) Normalized leakage rate. (b) Beam angle.

### A. Variations of $\alpha$ and $\beta$ Versus Frequency

Radiation characteristics of the structure at different frequencies must be obtained, so that the proper structure dimensions can be determined and from that the desired radiation pattern at the frequency of operation is realized.

The diagram in Fig. 3, which is created using the HFSS software and TRT, shows the changes in $\theta_m$ versus $\alpha/k_0$ at different frequencies. The proper values for the dimensions of the proposed structure are determined as $P = 32$ mm, $rw = 1.5$ mm, $sr = 11$ mm, $h = 0.8$ mm, and $a = 13$ mm and the results for different slot lengths are presented.

The dimensions of the structure, including $a$, $h$, and $P$, are chosen in a way that the cutoff frequency of the basic RSIW is 7.4 GHz; also n $= -1$ SH is generated and the scan angle starts from backward endfire. The proposed antenna also radiates in the broadside direction at the frequency of 9.5 GHz. However, based on (5), when $P$ changes the frequency at which the antenna radiates in, the broadside direction changes as well. According to the diagrams presented for $\theta_m$ and $\alpha/k_0$, the behavior of the proposed structure can be classified as follows.

1) In the first frequency range, which is the lowest frequency range and consists of the frequencies below the 7.4-GHz frequency, the dominant mode of the basic RSIW structure is not generated. In this frequency range, $\alpha$ has a very high value, but radiation does not take place and the input power is reflected back.

2) In the second frequency range, which consists of a frequency range from the 7.4 GHz cutoff frequency to the frequency of 9.5 GHz, the structure radiates from backward endfire to broadside. In this frequency range, $\theta_m$ varies from $-90°$ to $0°$. Also, as the frequency increases, $\alpha/k_0$ decreases to the point that when the main radiation beam is around broadside, it has a value of zero and the open stopband effect takes place at the maximum frequency in the second frequency range. When the open stopband appears around the broadside as a deep null, the n $= 0$ and n $= -2$ SHs have equal amplitudes with opposite directions, and the remaining SHs cancel each other in pairs in a similar manner as well.

3) The third frequency range starts when the radiation beam enters the forward quadrant and $\theta_m$ increases until the n $= -2$ SH or the next mode of the RSIW structure is generated. When the radiation beam reaches broadside, with increase in the frequency, the radiation beam has a smooth transition into the forward quadrant accompanied with a short pause. As mentioned before, at this point, $\alpha/k_0$ reaches a value of zero and after passing that, its value increases gradually.

4) The fourth frequency range starts from the frequency of 13.5 GHz at which the n $= -2$ SH is generated. Because two modes are generated at this frequency, there are two main lobes in the radiation pattern in different directions and the radiation pattern. By investigating different antenna parameters, the open stopband effect around the broadside can be improved. As shown in Fig. 3, different slot lengths are investigated, and it is clear that for different values of slot length, there is a different frequency at which $\alpha/k_0$ reaches its maximum value. As the slot is not acceptable any more, the length increases, and the frequency of maximum $\alpha/k_0$ decreases. The open stopband can be controlled by properly choosing the slot length. The maximum value of $\alpha/k_0$ can be placed at broadside at which $\alpha/k_0$ is usually zero in the conventional designs; consequently, the open stopband effect is suppressed. It can be observed in Fig. 3 that for $Ls = 27$ mm, maximum value for $\alpha/k_0$ takes place at a frequency corresponding to the broadside and the open stopband is eliminated. By comparing the results obtained from HFSS software and the TEN, it is observed that there is a good agreement between the two groups of the results and from that the high accuracy level of the TEN method can be concluded.

### B. Dependence of $\alpha$ and $\beta$ on Geometrical Parameters

Values of $P$, $Ls$, $h$, $rw$, and $sr$ influence $\alpha/k_0$ and $\beta/k_0$ values and changes in the former group of parameters result in changes in the latter group. Also, according to (8) and (9), variations in $\alpha/k_0$ and $\beta/k_0$ result in changes in bandwidth and radiation angle values, respectively. Consequently, for reaching the desired radiation pattern characteristics, five parameters including $P$, $Ls$, $h$, $rw$, and $sr$ have to be controlled.

Variations of $\alpha/k_0$ and $\beta/k_0$ at the frequency of 9.5 GHz, at which the open stopband occurs, have also been investigated for different parameters.

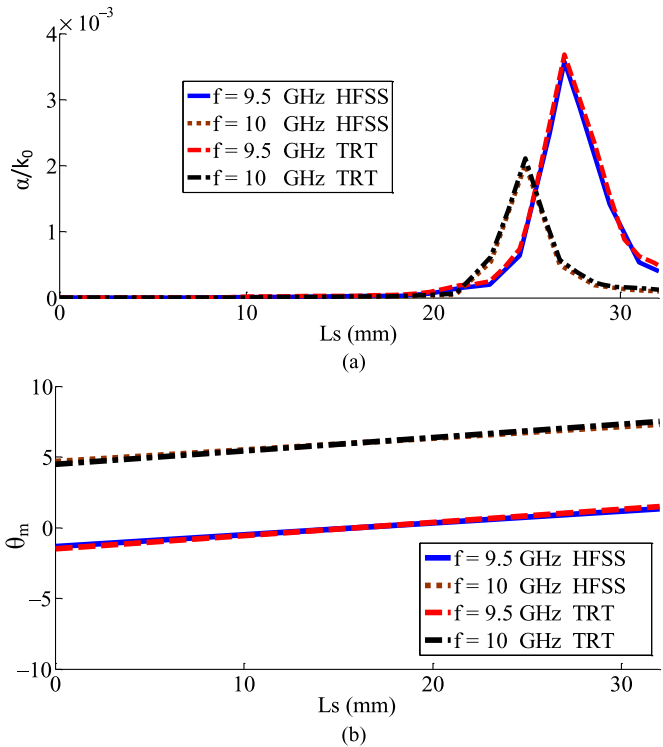

Fig. 4. Normalized leakage constant and beam angle for the proposed LWA versus the slot length at 9.5 and 10 GHz.

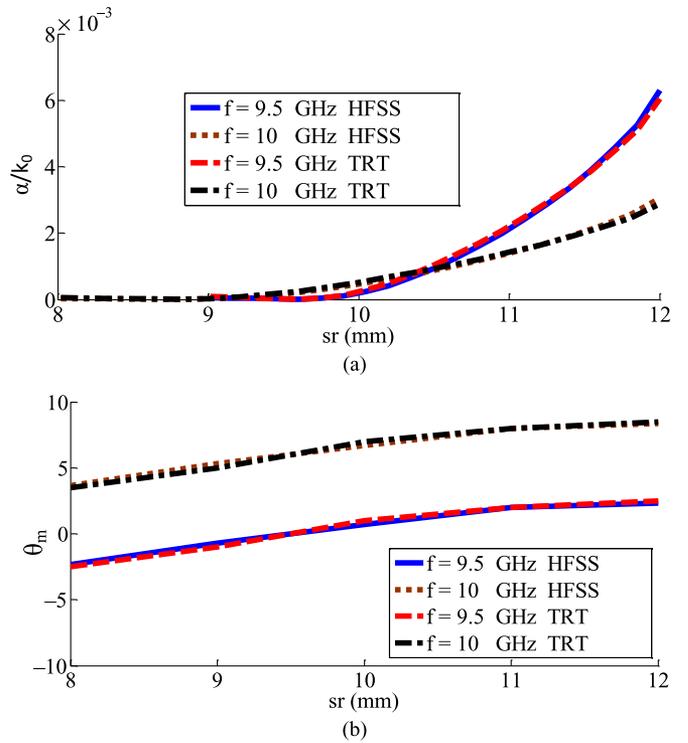

Fig. 5. Normalized leakage constant and beam angle for the proposed LWA versus the space of ridges at 9.5 and 10 GHz.

Fig. 4 depicts the changes in $\theta_m$ and $\alpha/k_0$ versus the changes in $Ls$ at the frequencies of 9.5 and 10 GHz. $\theta_m$ changes for about 2.5° with regard to the variations in $Ls$. When $Ls = 0$, i.e., when there is no discontinuity on the RSIW, $\alpha/k_0$ equals zero and there is no leakage. By increasing the slot length, $\alpha/k_0$ increases accordingly and reaches its maximum value when the slot lengths are 27 and 25 mm at the frequencies of 9.5 and 10 GHz, respectively.

Fig. 5 shows the effect of variations in $sr$ on $\alpha/k_0$ and $\theta_m$. When $sr$ is changed, the width of the first SIW layer changes and consequently $\beta/k_0$ of the structure and then $\theta_m$ are changed. By changing $sr$ from 8 to 12 mm at the frequencies of 9.5 and 10 GHz, the value of $\theta_m$ changes for 3.8° and 4.2°, respectively.

Fig. 6 shows the variations of $\theta_m$ and $\alpha/k_0$ when the widths of the ridges vary and $sr$ remains constant. In this figure, $sr$ equals $a - 2.5$ mm. As $a = 13$ mm, the sum of the widths of the ridges on the left and right sides of the slot, as shown in Fig. 1, has to be equal to 2.5 mm. When the width of the ridge on the left is equal to the width of the ridge on the right; in other words, when $rw = 1.25$ mm, the structure is symmetrical and $\alpha/k_0$ has a value of zero. As the difference between the widths of the two ridges increases, which means $rw$ becomes bigger or smaller than 1.25 mm, $\alpha/k_0$ increases because of the increase in the asymmetry of the structure. As $sr$ has a fixed value in this situation and the width of the lower waveguide in the RSIW structure remains almost fixed, $\beta/k_0$ is approximately fixed, and $\theta_m$ changes for about 0.5°. According to (5), $\beta_{-1}/k_0$ changes because of the changes in the period of the structure. Fig. 7 shows the effect of the changes in $P$ on

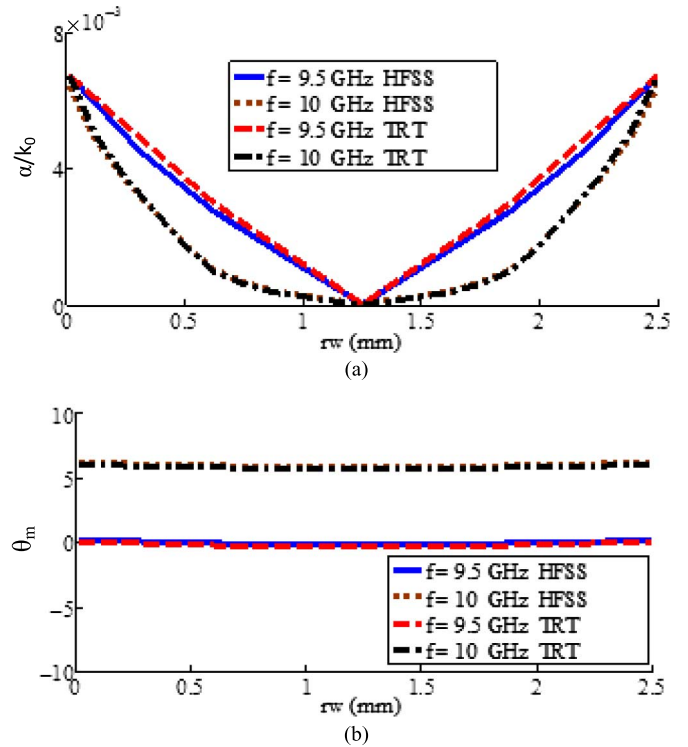

Fig. 6. Normalized leakage constant and beam angle for the proposed LWA versus the width of the ridge at 9.5 and 10 GHz.

$\alpha/k_0$ and $\beta/k_0$ at two different frequencies. When $P$ changes from 30 to 35 mm, $\theta_m$ changes from −13° to 25.6° and −8.5° to 32.5° at the frequencies of 9.5 and 10 GHz, respectively. In addition, the value of $\Delta\theta$ changes for about $10°$ at 9.5 GHz and $7°$ at 10 GHz. Although $\theta_m$ faces sharp changes, $\alpha/k_0$ changes slightly at both frequencies.

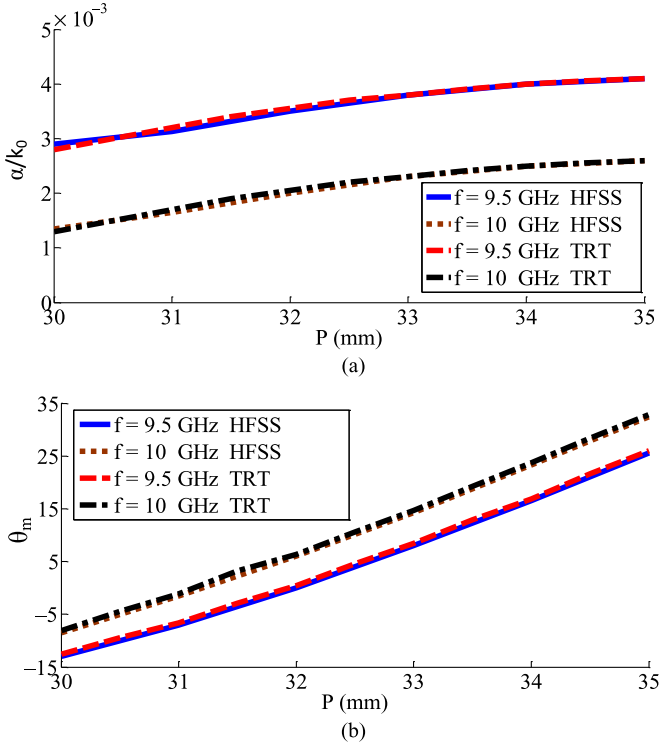

Fig. 7. Normalized leakage constant and beam angle for the proposed LWA versus the structure period at 9.5 and 10 GHz.

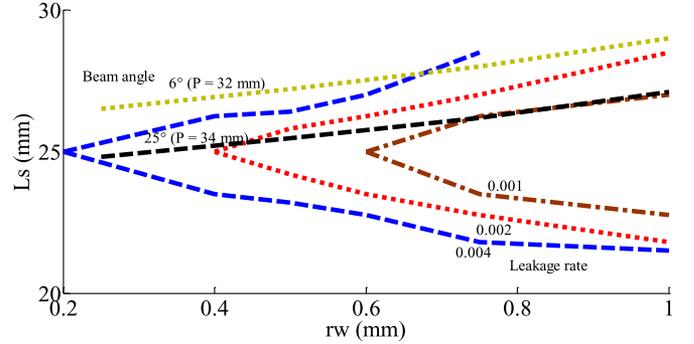

Fig. 8. Leaky-mode contour curve.

TABLE I
VALUES OF $rw$ AND $Ls$ FOR SLL $= -25$ dB AND $\theta_m = 6°$

| n | rw (mm) | Ls (mm) |
|---|---------|---------|
| 1 | 1.4     | 13.75   |
| 2 | 1.68    | 13.64   |
| 3 | 1.95    | 13.45   |
| 4 | 2.15    | 13.25   |
| 5 | 2.15    | 13.25   |
| 6 | 1.95    | 13.45   |
| 7 | 1.68    | 13.64   |
| 8 | 1.4     | 13.75   |

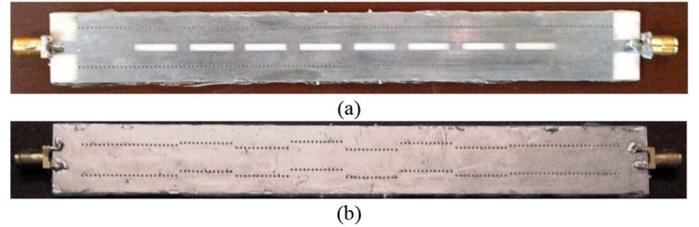

Fig. 9. Fabricated Antenna. (a) Top view. (b) Bottom view.

In the design of the proposed antenna, the height of the ridges $h$ is fixed. This is even as $\alpha$ can be easily controlled by changing $h$ in the structure; however, having different values for $h$ results in a multilayer structure with more complexity and higher manufacturing cost.

In all of the figures exhibiting diagrams of variations of different parameters obtained from the TRT and HFSS software, a good agreement between the results is observed, which confirms the preciseness of the TRT attained based on the accurate and multimode TEN. Figs. 4–7 show that $P$ is the most important parameter in changing the radiation beam angle and $rw$ and $Ls$ are the most effective parameters in controlling $\alpha/k_0$. In addition, Fig. 3 indicates that $Ls$ makes controlling the open stopband possible. However, other parameters affect the radiation pattern as well and need to have proper values.

## VI. DESIGN AND MEASUREMENT OF THE TAPERED LWA

For obtaining the desired radiation pattern, the pointing direction of all slots must be kept unchanged. In situations where a uniform array is designed, the slots have a similar $\theta_m$ and, i.e., because they all have similar parameter values. But radiation pattern as well and need to have proper values.

When $\alpha/k_0$ has to be tapered along the LWA structure for SLL optimization, each slot would have a different $\theta_m$, resulting in a wider beamwidth. As shown in Figs. 4–7, by changing each parameter in the geometry of the proposed antenna, $\alpha/k_0$ and $\theta_m$ are simultaneously changed. To keep $\beta/k_0$ fixed when $\alpha/k_0$ changes, some parameters need to be changed simultaneously.

A dispersion contour curve similar to those used in [23]–[25] is obtained and shown in Fig. 8. In this figure, $\alpha/k_0$ from zero to a given leakage rate can be obtained by changing $Ls$ and $rw$, even as $\theta_m$ is maintained constant. For producing this diagram, the results based on Figs. 4, 6, and 7 are used and different values of $\alpha/k_0$ and $\theta_m$ regarding the changes in $Ls$, $P$, and $rw$ are calculated. The mentioned diagram is shown in Fig. 8.

However, for considerably changing the radiation direction, $P$ needs to be changed. If the value of $P$ is fixed, similar to the situation in Fig. 8, and other parameters are changed, the radiation angle is almost equal to the angle corresponding to that particular $P$. For realizing the desired SLL, electric field distributions in the structure has to be controlled through using distributions such as the Chebyshev distribution.

The variations of the leaky mode leakage rate of each of the slots can be computed using the illumination function expressed by the following equation:

$$\alpha(N) = \frac{P_N}{1/\eta \sum_{n=1}^{m} P_n - \sum_{n=1}^{N-1} P_n} \quad (10)$$

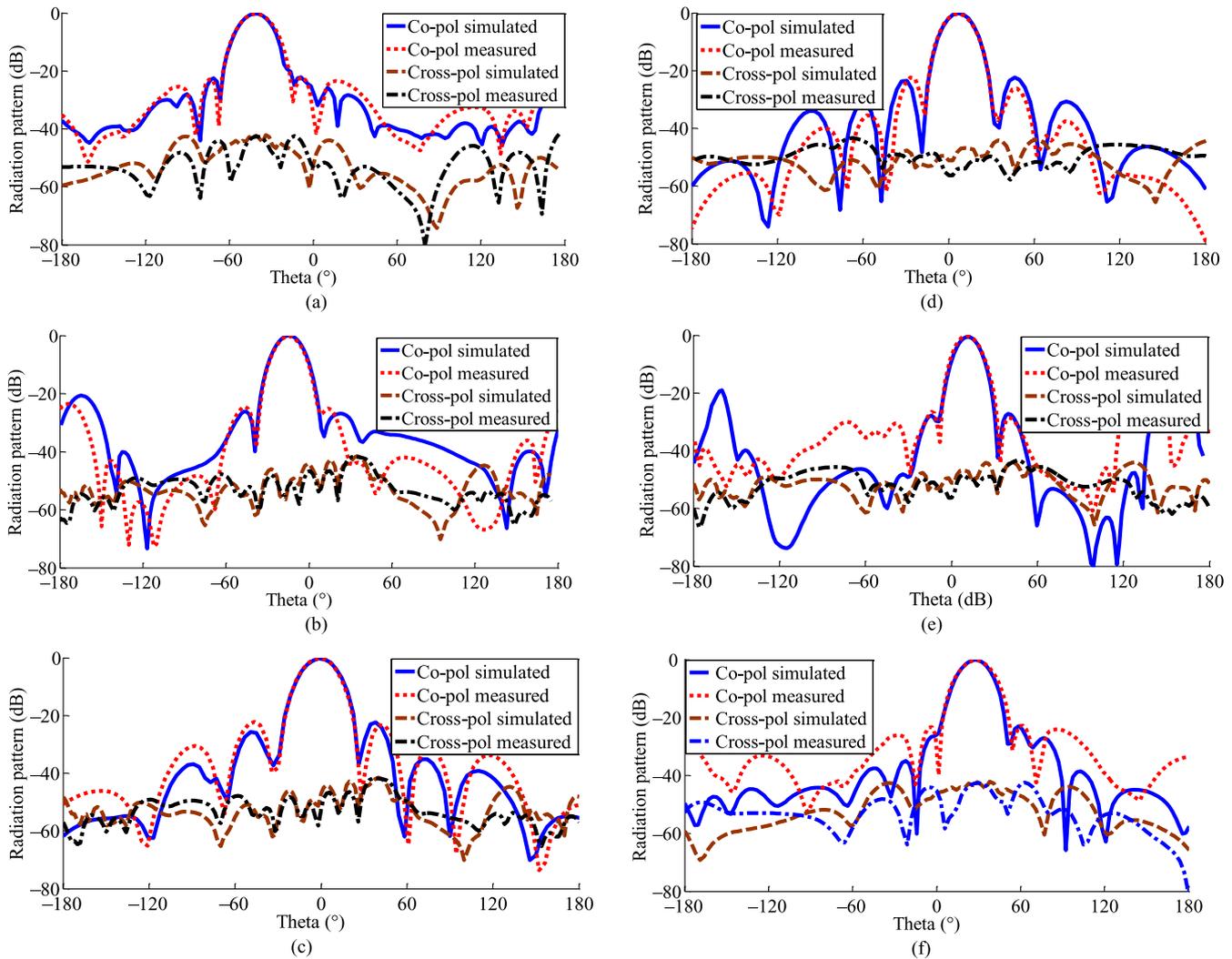

Fig. 10. Normalized co- and cross-polarization radiation patterns for (a) 8, (b) 9, (c) 9.5, (d) 10, (e) 10.5, and (f) 12 GHz.

where $\eta$ is the antenna efficiency and is expressed as the ratio between the radiated power and the antenna input power. It might be necessary to note that about 5% of the input power reaches port 2 at different frequencies. $P_n$ is the normalized power radiated from the nth slot. Also, $m$ is the number of slots. Table I shows the changes in $rw$ and $Ls$ at the frequency of 10 GHz for realizing an SLL of $-25$ dB. Using the dispersive curve shown in Fig. 8, $rw$ and $Ls$ dimensions are chosen in such a manner that the expected leakage rate for the tapering in the structure is realized and $\theta_m$ has a fixed value of $6°$. The proposed structure was fabricated and tested based on the calculated dimensions.

As it can be seen in Fig. 9, for each substrate, a Rogers4003 substrate with $\varepsilon_r = 3.55$ and $h = 0.8$ mm is used for manufacturing the proposed RSIW slot array LWA. Fig. 10(a)–(f) plots and compares the normalized radiation patterns of the measured and simulated co-polarization and cross-polarization at the frequencies of 8, 9, 9.5, 10, 10.5, and 12 GHz, respectively. In this frequency range, the radiation beam scans from $-35°$ to $35°$.

Radiation patterns resulted from simulations and measurements show good agreement. It is shown that for various scanning angles, SLL reduction is realized. Radiation efficiency and gain values of the proposed structure at the frequencies of 8, 9, 9.5, 10, 10.5, and 12 GHz are shown in Table II. The proposed antenna has a gain of about 11 dB and the minimum radiation efficiency is about 91%.

Fig. 11 shows the values of the scattering parameters of the designed and tapered LWA. Measured and simulated S11 parameters are in good agreement. In the frequency range of 8–12 GHz, the measured reflections of the proposed structure are approximately less than $-10$ dB.

TABLE II
GAIN AND RADIATION EFFICIENCY OF THE PROPOSED SLOTTED LWA

| Frequency (GHz) | Gain (dB) | Efficiency (%) |
|---|---|---|
| 8 | 10.5 | 90 |
| 9 | 10.9 | 91 |
| 9.5 | 11 | 90.5 |
| 10 | 11.5 | 91.2 |
| 10.5 | 11.8 | 91.4 |
| 12 | 12.5 | 91 |

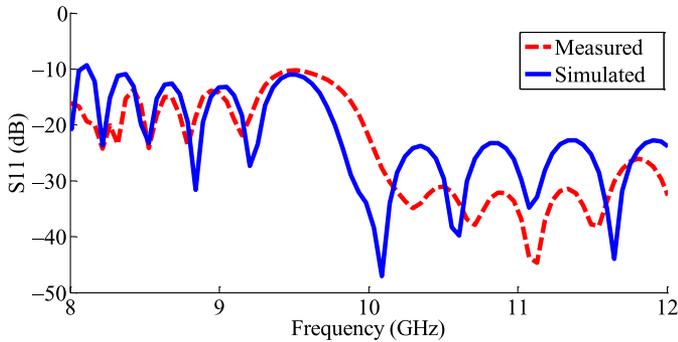

Fig. 11. Measured and simulated of S11 for proposed PLWA.

## VII. CONCLUSION

A collinear-slotted RSIW PLWA is proposed. As the slots have no offset from the centerline of the upper plane of the RSIW, the radiation pattern exhibits a low cross-polarization and there is no second-order beam. Through controlling different parameters of the structure, an antenna is proposed that is capable of scanning from backward endfire into the forward quadrant by changing the frequency. In addition, the open stopband in the broadside is controlled. To avoid beam broadening in the proposed antenna because of tapering the field distribution of the antenna aperture, by simultaneously changing the slots lengths and widths of the ridges, a fixed $\alpha$ and constant $\beta$ are achieved. The proper selection of the period of the ridges resulted in an increase in the maximum scan angle. The proposed structure was simulated using HFSS software and then manufactured. The measured SLL and cross-polarization have values of $-25$ and $-45$ dB, respectively, which show good agreement with simulation results.

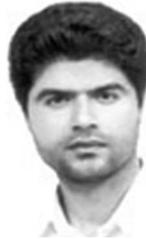

**Alireza Mallahzadeh** (M'12–SM'15) received the B.S. degree in electrical engineering from Isfahan University of Technology, Isfahan, Iran, in 1999, and the M.S. and Ph.D. degrees in electrical engineering from Iran University of Science and Technology, Tehran, Iran, in 2001 and 2006, respectively.

He is a Member of Academic Staff, Faculty of Engineering, Shahed University, Tehran, Iran. He has participated in many projects relative to antenna design, which resulted in fabricating different types of antennas for various companies. His research interests include numerical modeling and microwaves.

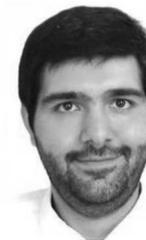

**Sajad Mohammad-Ali-Nezhad** (M'15) received the B.Sc. degree in electronic engineering from Shahid Chamran University, Ahwaz, Iran, in , and the M.Sc. and Ph.D. degrees in communication engineering from Shahed University, Tehran, Iran, in 2010 and 2015, respectively.

Currently, he is the Head with the Department of Electrical and Electronics Engineering, University of Qom, Qom, Iran. His research interests include leaky wave antennas, printed circuit antennas, array antennas, phased array antennas, MIMO antennas, RFID tag antennas, frequency selective surface, electromagnetic compatibility, microwave filters, and electromagnetic theory.